\documentclass[longauth]{aa} % for the long lists of affiliations NORMAL ONE TO USE
\usepackage{graphicx}
\usepackage{txfonts}
\usepackage{natbib}
\bibpunct{(}{)}{;}{a}{}{,}

\usepackage{txfonts}
\begin{document} 
   \title{The Gaia-ESO Survey: Stellar radii in the young open clusters NGC~2264, NGC~2547 and NGC~2516
\thanks{Based on observations collected with the FLAMES spectrograph at VLT/UT2 
telescope (Paranal Observatory, ESO, Chile), for the Gaia- ESO Large Public Survey (188.B-3002).}}
%
%   \subtitle{}

\author{R.~J.~Jackson \inst{1}
\and
R.~D.~Jeffries \inst{1}
\and
S. Randich \inst{2}
\and
A.~Bragaglia \inst{3}
\and
G.~Carraro \inst{4}
\and
M.~T.~Costado \inst{5}
\and
E.~Flaccomio \inst{6}
\and
A.~C.~Lanzafame \inst{7}
\and
C.~Lardo \inst{8}
\and
L.~Monaco \inst{9}
\and
L.~Morbidelli \inst{2}
\and
R. Smiljanic \inst{10}
\and
S. Zaggia \inst{11}
}

\institute{Astrophysics Group, Keele University, Keele, 
      Staffordshire ST5 5BG, UK\\
\email{r.j.jackson@keele.ac.uk}
\and
INAF - Osservatorio Astrofisico di Arcetri, Largo E. Fermi 5, 50125,
Florence, Italy
\and
INAF - Osservatorio Astronomico di Bologna, via Ranzani 1, 40127,
Bologna, Italy
\and
European Southern Observatory, Alonso de Cordova 3107 Vitacura,
Santiago de Chile, Chile
\and
Instituto de Astrof\'{i}sica de Andaluc\'{i}a-CSIC, Apdo. 3004, 18080,
Granada, Spain
\and
INAF - Osservatorio Astronomico di Palermo, Piazza del Parlamento 1,
90134, Palermo, Italy
\and
Dipartimento di Fisica e Astronomia, Sezione AstrofiFsica,
Universit\`{a} di Catania, via S. Sofia 78, 95123, Catania, Italy
\and
Astrophysics Research Institute, Liverpool John Moores University, 146
Brownlow Hill, Liverpool L3 5RF, United Kingdom
\and
Departamento de Ciencias F\'isicas, Universidad Andr\'es Bello,
Rep\'ublica 220, 837-0134 Santiago, Chile
\and
Department for Astrophysics, Nicolaus Copernicus Astronomical Center,
ul. Rabia\'{n}ska 8, 87-100 Toru\'{n}, Poland
\and
INAF - Padova Observatory, Vicolo dell'Osservatorio 5, 35122 Padova,
Italy
}

\date{}

% \abstract{}{}{}{}{} 
% 5 {} token are mandatory
 
  \abstract
   {Rapidly rotating, low-mass members of eclipsing binary systems have
   measured radii that are significantly larger than predicted by standard
   evolutionary models. It has been proposed that magnetic activity is
   responsible for this radius inflation.}
   {By estimating the radii of low-mass
     stars in three young clusters (NGC\,2264, NGC\,2547, NGC\,2516,
     with ages of $\sim$5, $\sim$35 and $\sim$140\,Myr respectively), we aim to
   establish whether similar radius inflation is seen in single,
   magnetically active stars.}
   {We use radial velocities from the Gaia-ESO Survey (GES) 
  and published photometry to establish cluster membership and then combine GES
     measurements of projected equatorial velocities with published
     rotation periods to estimate the average radii for groups of
     fast-rotating cluster members as a function of their luminosity and
     age. The average radii are compared with the predictions of both
     standard evolutionary models and variants that include magnetic
     inhibition of convection and starspots.}
   {At a given luminosity, the stellar radii in NGC~2516 and NGC~2547 
    are larger than predicted
     by standard evolutionary models at the ages of these clusters. The
     discrepancy is least pronounced and not significant ($\simeq 10$ per cent) in ZAMS stars
     with radiative cores, but more significant in lower-mass, fully convective pre
     main-sequence cluster members, reaching $\simeq 30\pm 10$ per
     cent.  The uncertain age and distance of NGC~2264 preclude a
     reliable determination of any discrepancy for its members.}
   {The median radii we have estimated for low-mass fully convective
     stars in the older clusters are 
     inconsistent (at the $2-3 \sigma$ level) with non-magnetic
     evolutionary models and more consistent with
     models that incorporate the effects of magnetic
     fields or dark starspots. The available models suggest this
     requires either surface magnetic fields exceeding 2.5 kG, spots that
     block about 30 per cent of the photospheric flux, or a more moderate
     combination of both.}

\titlerunning{The Gaia-ESO survey: Stellar radii in young open clusters}
\authorrunning{R. J. Jackson et al.}

   \maketitle

\section{Introduction}

\subsection{The radii of low-mass stars}

The radius of a star is one of its most fundamental properties, that
ought to be correctly predicted by stellar models. Yet precise
measurements of K- and M-dwarfs in eclipsing binary systems have shown
that, for a given mass, the radii of stars with
$0.2<M/M_{\odot}<0.8$ are 10--15 per cent larger than predicted
by current evolutionary models. Hence, for a given luminosity, the
effective temperature can be underestimated by up to 7.5 per cent
(e.g. Lopez-Morales 2007, Morales et al. 2009, Torres et
al. 2010). 

Stars in these short-period, tidally locked binaries are fast-rotating
and highly magnetically active; a working hypothesis is that the larger
radii are caused either by the suppression of convection by interior,
dynamo-generated magnetic fields (e.g. Mullan \& MacDonald 2001;
Chabrier, Gallardo \& Baraffe 2007; Feiden \& Chaboyer 2012, 2013), or by the
blocking of flux at the surface by cool, magnetic starspots (Chabrier
et al. 2007, Macdonald \& Mullan 2012, Jackson \& Jeffries 2014b).
Similarly high levels of rotation and magnetic activity are also a characteristic of
{\it young}, low-mass stars. 
If rotationally induced magnetic fields or starspots do affect the
radii of fast rotating stars, then this could significantly
alter the masses and ages inferred for such objects from their locations in
the Hertzprung-Russell diagram, change estimates of the radii and
hence densities of any transiting exoplanets that are discovered around
them and alter the age-dependence of lithium depletion in their
photospheres (Jackson \& Jeffries 2014a; Somers \& Pinsonneault 2014, 2015).

A suspicion remains however, that the oversized radii could be directly
connected with the close binary nature of the eclipsing stars for which
precise radii and masses are available. Tidal synchronisation could
modify internal rotation profiles, convection patterns and magnetic
activity. If an inflated radius really is a consequence of
rotation/magnetic activity and not just binarity, then a simple test is
that rapidly rotating, {\it single} K- and M-dwarfs should also be
bigger than both inactive stars and the model predictions. Although the
masses of single stars are inaccessible, it is possible to measure
their radii as a function luminosity. Interferometric techniques have
been used to determine radii for nearby K- and M-dwarfs with an
estimated mass range of $0.3<M/M_{\odot}<0.8$ (Boyajian et al. 2012).
Unfortunately, almost all these nearby stars are magnetically {\it
  inactive}, so the radius-luminosity relation that can be derived is
applicable to relatively old, slow-rotating, main sequence (MS)
stars. This relationship shows satisfactory agreement with the
predicted radii of evolutionary models (e.g. Baraffe et al. 1998; 
Dotter et al. 2008) for K- and early M-dwarfs, but later M-dwarfs
(spectral types beyond M2) show lower effective temperatures than 
predicted by evolutionary models (Veeder 1974), with current models 
underestimating radii by $\sim 5$ per cent (Boyajian et al. 2012).

Extending this comparison to magnetically active, single, low-mass
stars is difficult since there are none close enough to allow precise
interferometric radius measurements. Instead Jackson, Jeffries \&
Maxted (2009) and Jackson \& Jeffries (2014b) have used the product of
individual rotation periods ($P$) and projected equatorial velocities
($v \sin i$) to estimate projected stellar radii ($R \sin i$). Assuming
random axial orientations, such measurements can provide average radii
for groups of stars. The technique has been applied as a function of
luminosity in two young open clusters (NGC~2516 and the
Pleiades). Jackson \& Jeffries (2014b) found that the highly
magnetically active K- and M-dwarfs in these clusters (aged
$\simeq$130\,Myr - Meynet, Mermilliod and Maeder 1993) showed
significant radius inflation relative to the empirical locus defined by
inactive MS field stars; the mean increase in radius, at a given
luminosity, ranges from $13\pm3$ per cent for MS K-dwarfs to $40\pm4$
per cent for the lower luminosity pre main-sequence (PMS) M-dwarfs that
are still descending their fully convective Hayashi tracks.

\subsection{The Gaia-ESO Survey}

The Gaia-ESO survey (GES) is employing the FLAMES multi-object
spectrograph (Pasquini et al. 2002) on the VLT UT-2 (Kueyen) telescope
to obtain high quality, uniformly calibrated spectroscopy of $>10^5$
stars in the Milky Way over a 5 year period (Gilmore et al. 2012; Randich \& Gilmore
2013). The survey, which began at the end of 2011, 
includes stars from the halo, bulge, thick and thin
discs, as well as in star forming regions and clusters of all
ages. Samples are chosen from photometric surveys with the aim of
characterizing the chemical and kinematic evolution of these
populations. Analysis of these spectra will provide a rich dataset of
chemical and dynamical parameters which, when combined with proper
motions and parallaxes from the Gaia satellite, will provide full
three-dimensional velocities and chemistry for a large and representative
sample of stars. 

In addition to radial velocity ($RV$) data, the GES measures the
$v\sin i$ of target stars (Koposov et
al. in preparation). In the case of young clusters, $RV$ can be used to
confirm cluster membership and $v\sin i$ data can be combined with
published measurements of rotation periods to estimate the $R\sin i$
values for members of the cluster (e.g. Baxter et al. 2009, Jackson et al. 2009). 
As the GES progresses the opportunity arises
to estimate the radii of both MS and PMS stars for a number of clusters,
where rotation period data are available, using uniformly derived and calibrated values of $v\sin
i$. By determining the radii of low-mass stars as a function of their
luminosity in clusters of different ages, we can test evolutionary
models, search for the signatures of radius inflation by magnetic
activity and investigate if there is any dependence on whether
stars have reached the zero age main sequence (ZAMS) (e.g. see Jackson \& Jeffries 2014b).

In this paper we use GES data to estimate the average radii for stars
in three young clusters -- NGC~2516, NGC~2547 and NGC~2264. In section
2 we describe the clusters and our database and use the $RV$ of
spectroscopic targets to assign cluster membership.  In Sect.~3 we
use the $v\sin i$ values of confirmed cluster members with known
periods to determine average radii as a function of luminosity. Finally,
in Sect.~4 we compare the radius data for each of the three clusters
with the radii as a function of luminosity and age predicted by
different evolutionary models, some of which include the effects of
magnetic fields and starspots.

\section{Cluster members and their projected rotation velocities}

\subsection{Cluster properties}

NGC~2516 is the oldest of the three clusters considered in this
paper. Meynet, Mermilliod \& Maeder (1993) give an age of
141\,Myr. This is consistent with more recent estimates of age of
$\simeq 150$\,Myr from lithium depletion in low mass stars (Jeffries,
James \& Thurston 1998) and $125\pm 25$\,Myr from the nuclear turn off for high
mass stars (Lyra et al. 2006). We adopt the intrinsic distance modulus of
$7.93\pm 0.14$ mag and cluster reddening of $E(B-V)=0.12 \pm0.02$ mag given by
Terndrup et al. (2002) based on main sequence fitting.

NGC~2547 is a younger cluster and our dataset may be
expected to contain both PMS and ZAMS stars . We adopt an age of
$35 \pm 3$\,Myr based on lithium depletion (Jeffries \& Oliveira
2005). This is consistent with the age of 38.5$^{+3.5}_{-6.5}$\,Myr
found from main sequence fitting by Naylor \& Jeffries (2006), who also
give an intrinsic distance modulus of $7.79^{+0.11}_{-0.05}$ mag and reddening
of $E(B-V)=0.12\pm 0.05$ mag.

NGC~2264 is the youngest cluster considered here with an estimated age
of 2--7\,Myr depending, amongst other things, on the adopted
distance to the cluster, which varies between 400\,pc (Dzib et al. 2014)
and 913\,pc (Baxter et al. 2009). In this paper we adopt the distance
$777\pm 12$\,pc given by Turner (2012) based on model isochrone
fitting. This is consistent with a distance of $760\pm 49$\,pc from fitting
the ZAMS (Sung et al. 1997) and the parallax-based
distance of $738 \pm 45$\,pc obtained from two maser sources thought to be
located in NGC~2264 (Kamezaki et al. 2013). Having adopted the
distance given by Turner (2012) we also adopt their average reddening of
$E(B-V)= 0.075 \pm 0.003$\,mag and age of $\sim 5.5$\,Myr. We note that at best
this age represents a median value; Sung and Bessell (2010) estimate
an age spread of 2--3\,Myr about the median age.

All three clusters show near solar-metallicity, with [Fe/H] values,
consistently determined from GES spectroscopy, that are within $\pm
0.1$ dex of solar (Magrini \& Randich 2015).

\begin{table*}
	\caption{ Photometric and spectroscopic data of GES Survey targets in clusters in NGC\,2547, NGC\,2516, NGC\,22264 downloaded from the Edinburugh GES archive. The full
	Table~1 is available at the CDS.}
    \begin{tabular}{lllllllllllll}\hline\hline

Cluster & Object	&	I$_{\rm C}$	&	$J_{\rm 2MASS}$	&	$H_{\rm 2MASS}$	&	$K_{\rm 2MASS}$	&	$S/N$	&	$RV$	&	$v\sin i$	&	$\log T_{\rm eff}$	&	No.	&	$\Delta_{RV}$	&	$\Delta_{v\sin i}$	\\		
 & 	&	mag.	& mag.	&	mag.	&	mag.	& &	km\,s$^{-1}$	&	km\,s$^{-1}$	&	 & Spec	&	km\,s$^{-1}$	&		\\\hline
	
NGC2264 &06392497+0933151	&	15.26	&	13.84	&	13.28	&	13.06	&	8.72	&	31.10		&	4.3	&	3.552	&	2	&	0.54	&	0.66	\\
NGC2264 &06392506+0942515	&	14.26	&	13.21	&	12.48	&	12.36	&	15.02	&	20.08		&	1.1	&	3.560	&	2	&	0.36	&	1.43	\\
NGC2264 &06392535+0943147	&	13.74	&	12.05	&	11.19	&	10.90	&	19.39	&	-7.64		&	1.6	&	3.712	&	2	&	0.36	&	4.31	\\
NGC2264 &06392550+0931394	&	15.32	&	13.77	&	12.86	&	12.38	&	27.40	&	17.98		&	4.7	&	3.544	&	2	&	0.29	&	0.23	\\
NGC2264 &06392649+0943298	&	15.56	&	13.94	&	13.26	&	12.98	&	23.86	&	15.30		&	19.8&	3.770	&	2	&	0.63	&	0.09	\\
NGC2264 &06393939+0945215	&	13.97	&	12.90	&	12.04	&	11.54	& 55.61	&	20.18		&	11.8&	3.621	&	4	&	0.21	&	0.05	\\
NGC2264 &06393961+0945442	&	13.03	&	11.98	&	11.37	&	11.18	&	52.77	&	43.60		&	1.6	&	3.610	&	4	&	0.22	&	1.22	\\\hline
    \multicolumn {12} {l} {No. Spec indicates the number of separate observations co-added to produce the target spectrum}\\
    \multicolumn {12} {l} {$\Delta_{RV}$ and $\Delta_{v\sin i}$ are
      calculated values of the  absolute precision in $RV$ and fractional precision in $v\sin i$, as described in Sect. 2.2}\\
     
  \end{tabular}
\label{tab1}
\end{table*}

\begin{table}
 \centering
	\caption{Cluster properties and target numbers}
		\begin{tabular}{lccc}\hline\hline
                              &NGC\,2516   & NGC\,2547   &NGC\,2264\\\hline
Age (Myr)              & 141              &    35               &   5.5\\
$(m-M)_0$               & 7.93            &    7.79               &   9.45\\
E(B-V) (mag)                 & 0.12              &    0.12               &   0.075\\
$A_K$ (mag)                &      0.04         &       0.04          &    0.02     \\
No. of targets observed&  743  &  450  & 1707 \\     
No. potential members (a)       &618              &376                     &1525\\
No. members (b)   &459                &156                     &350 (604)\\ 
Targets with Known period      &77                  &84                      &226\\
Members with $R\sin i$ (c)        &32                  &45                    &90 (157)\\\hline

		 \multicolumn {4} {l} {(a) With $S/N>5$, valid I-mag, K-mag and $\log g>3.5$ (if known)}\\
		 \multicolumn {4} {l} {(b) Based on the probability of
                   membership from the $RV$ being $>0.8$}\\
		 \multicolumn {4} {l} {(c) Member with known period, $v\sin i >5$\,km\,s$^{-1}$ and $\Delta _{v\sin i}<0.2$}\\
		 \multicolumn {4} {l} {Nos. in brackets are for relaxed membership criteria, see Sect.~2.4}\\
		 		 \end{tabular}
	\label{tab:tab2}
\end{table}

\begin{figure*}
\centering
\includegraphics [width = 180mm]{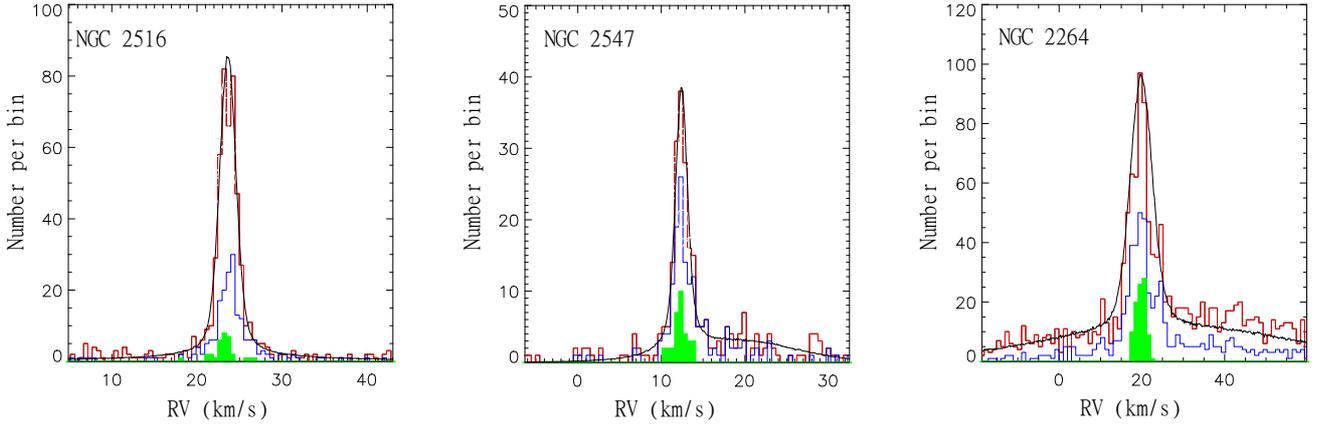}
\caption{The number density of GES cluster targets in open clusters
  NGC\,2516, NGC\,2547 and NGC\,2264 as a function of $RV$. the upper (red)
  histograms show all potential members with $S/N >5$. The lower
  (blue) histograms show the number of faster rotators ($v\sin
  i > 5$\,km\,s$^{-1}$) and the solid (green) histograms show stars
  identified as cluster members with measured rotation periods (see section
  3.1). The black curve shows the probability density fit to the
  measured data comprising of the sum of two quasi-Gaussian distributions; the first,
  a narrower distribution, represents cluster members, and a second much
  broader distribution represents non-members.}
\label{fig1}
\end{figure*}

\begin{figure*}
\centering
\includegraphics [width = 180mm]{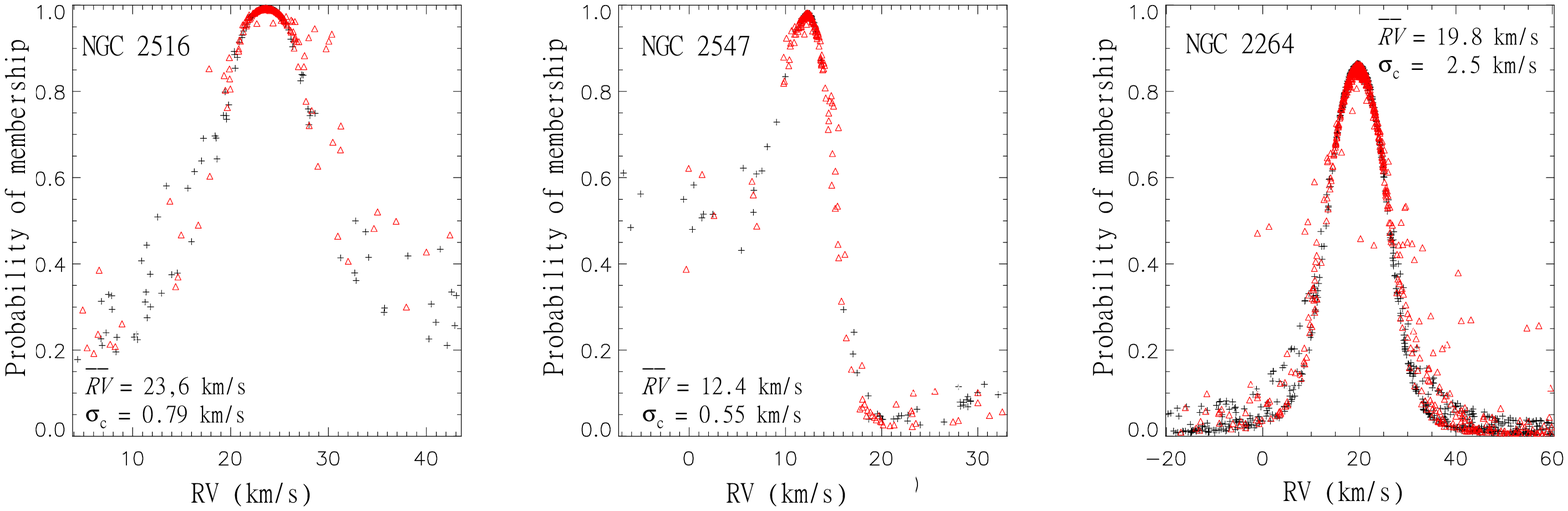}
\caption{The probability of cluster membership as a function of $RV$
  for open clusters NGC\,2516, NGC\,2547 and NGC\,2264 derived from the
  maximum likelihood fit of a double quasi-Gaussian distribution to the
  data in Fig.~1. Black crosses show probabilities for slower
  rotators ($v\sin i \leq 5$ km\,s$^{-1}$), red triangles show
  probabilities for faster rotators. Estimated values of the mean velocity, $\overline{RV}$ and
  dispersion $\sigma_c$ of the cluster population are shown on the plots.}
\label{fig2}
\end{figure*}

\subsection{Cluster targets with resolved projected equatorial velocities}

Targets in these clusters were observed at the VLT on a number of dates
between April 2012 and December 2013. The targets were selected using a
combination of optical and infrared photometry and chosen
using very broad regions of colour-magnitude space, so as to target all possible cluster members (or at
least make them available for fibre placement -- see Bragaglia et
al. in preparation). 
The raw spectroscopic data were processed with
pipelines developed at the Cambridge Astronomical Survey Unit, to
produce wavelength-calibrated, sky-subtracted spectra and estimates of
$RV$ and $v \sin i$  (Lewis et al; Koposov et al. in
preparation). These were passed to spectroscopic analysis working
groups (e.g. Lanzafame et al. 2015), which estimated astrophysical
parameters ($\log g$, $T_{\rm eff}$) and chemical abundances for a
subset of the targets with high quality spectra. The resulting data
and parameters were placed in the GES archive (and made available to
the GES consortium) at the Wide Field
Astronomy Unit at Edinburgh University\footnote{http://ges/roe.ac.uk/}. 
Data used in this paper is taken from tables 
Target, Spectrum, RecommendedAstroAnalysis and SpectrumNightly of the iDR2/iDR3 
release of GES data.
 
All the relevant photometry and spectroscopic data for GES survey
targets in NGC\,2516, NGC\,2547 and NGC\,2264, were taken from the
Edinburgh GES archive. 
The database contains $RV$ and $v\sin i$ values
derived from the analysis of both individual spectra (termed nightly
spectra) and combined spectra -- where all available nightly spectra are
co-added to produce a single target spectrum. All the results reported
here are given for combined spectra observed with the intermediate
resolution GIRAFFE spectrograph, using order sorting filter HR15N which 
gives a spectral wavelength range of 6445--6815$\AA$ at a resolution of 17000. The
targets lie in the brightness range of approximately $11<V<19$ for all
three clusters. The optical photometry originates from Jeffries, Thurston \&
Hambly (2001) and Irwin et al. (2007) for NGC\,2516; from Naylor et
al. (2002), Jeffries et al. (2004) and Irwin et al. (2008) for
NGC\,2547; and from Sung et al. (2008) for NGC\,2264. 2MASS near infrared
photometry is available for all targets (Skrutskie et al. 2006).

The uncertainties in $RV$ and $v\sin i$ were estimated, as a function
of temperature, $v \sin i$ and $S/N$, using the scaling parameters
$\Delta_{RV}$ and $\Delta_{v\sin i}$, the prescription defined by
Jackson et al. (2015)\footnote{Note that $\Delta_{RV}$ is expressed as
  an absolute uncertainty in km\,s$^{-1}$, whereas $\Delta_{v \sin i}$
  is a dimensionless fractional precision -- see Jackson et al. (2015)
  for details.}, and with the constants shown in table 3 of their paper
that are appropriate for order sorting filter HR15N. These scaling
parameters approximate to the standard deviations for a normal
distribution, but Jackson et al. showed that the underlying uncertainty
distributions have extended tails and are better represented by Student's
t-distributions with $\nu$ degrees of freedom, where $\nu =6$ for $RV$
and $\nu=2$ for $v\sin i$.  The intrinsic spectral resolution of the
data means that there is an effective resolution limit to $v \sin i$,
such that only objects with $v \sin i > 5$ km\,s$^{-1}$ have reliably
detected rotation. Data downloaded from the Edinburgh GES archive including 
calculated values of the $RV$ and $v\sin i$ precision for each target are listed in
Table 1, the full version of which is available at CDS. 

Targets for each cluster were selected initially from the database as
having both valid $I$ and $K$-magnitudes and a signal to noise ratio of the
combined spectra $S/N>5$. Where possible, a further selection was made
on the basis of surface gravity to remove field red giants (with $\log
g<3.5$) from the sample. Estimates of $\log g$ were available from the
Working Group (WG) analyses for 99 per cent of targets in NGC\,2516 but
only $\sim 15$ per cent of targets in the other two clusters. The
numbers of potential targets are shown in Table 2.

\subsection{Target mass and luminosity}
2MASS magnitudes of the targets in Table 1 were transformed to the CIT 
system using relations given by Carpenter (2001) then scaled
to absolute magnitudes using the distance moduli and extinctions
shown in Table 2 (assuming that the targets are members of the
cluster). Luminosities of targets in NGC\,2547 and NGC\,2516 were estimated 
from target $M_K$ values using the bolometic correction as a function of $(I-K)$ colour 
given by BHAC15  model isochrones (Baraffe et al. 2015) interpolated to
the cluster age. 
Masses were estimated from luminosities using the same models.  Luminosities in NGC\,2264
were estimated from $M_J$ values using the bolometic correction 
as a function of $(I-J)$ colour in order to minimise possible effects of any infra-red excess
due to circumstellar discs which may still be present in this relatively young cluster.

\subsection{Probability of membership}
Figure 1 shows histograms of the $RV$ of potential members for each clusters. 
Results are shown separately for (a) all potential members and (b) those with a $v\sin i
>5$\,km\,s$^{-1}$. The latter group of relatively rapidly rotating stars have a
higher probability of cluster membership, since older field stars are
more likely to be slower rotators. 

A maximum likelihood method 
was used to determine the probability of
cluster membership as a function of $RV$ by fitting a pair of
quasi-Gaussian distributions to the data for each star. The first
distribution defines cluster members, the second much broader
distribution represents the background population. Each probability
density function is the convolution of a Gaussian profile with the
distribution of measurement uncertainty incorporating the combined
effects of measurement precision and the projected
orbital velocities of an estimated fraction of binary stars (see
Cottaar, Meyer \& Parker 2012; Jeffries et al. 2014).

For the present calculation.
\begin{itemize}
\item the measurement uncertainty in $RV$ of individual targets is
  defined as a Student-t distribution with $\nu$=6 scaled according to
  the scaling constant for measurement precision $\Delta_{RV}$ (see Jackson
  et al. 2015).
\item Binary periods are assumed to follow the log normal distribution described by
  Raghavan et al. (2010) with $<\log P>=5.03$ and $\sigma_{\log
    P}=2.28$, 
where $P$ is the period in days.
\item The binary fraction varies with mass as $f_b = 0.27 +
  0.47(M/M_{\odot})$ over the range $0.2 < M/M_{\odot} < 1.2$.
  This relationship, estimated from the binary fraction as a function
  of mass given by Jeffries et al. (2001) for
  NGC\,2516, is assumed to apply to all three clusters here.
\item	The binary mass ratio is represented by a flat distribution
  between 0.1 and 1 and the eccentricity by a uniform probability
  density between 0 and 0.8 (Raghavan et al. 2010).
\item	Targets outside the range $\pm$20~km\,s$^{-1}$ of the cluster
  central velocity are assumed to be non-members and are excluded from
  the calculation. A range of $\pm$40\,km\,s$^{-1}$ is adopted for
  NGC\,2264, which appears to have a much broader intrinsic distribution.
\end{itemize}

Figure~1 shows the maximum likelihood fit of the number density of
targets as a function of $RV$ for each cluster. Figure 2 shows the
probability of membership as a function of $RV$ for each star, which is
calculated from the ratio of likelihoods that a target belongs to the
cluster or field populations. This depends strongly on $RV$ and more
weakly on the $RV$ measurement uncertainty. Note that the intrinsic
dispersions of the clusters quoted in Fig.~1 are upper limits to the
true dispersion. There is some evidence for mass-dependence in the
dispersion and for a small (probably unphysical) mass-dependent systematic drift in
$RV$ which inflates the dispersion seen in a group covering a wide mass
range. This is more carefully analysed in Jeffries et al. (in
preparation), however the accuracy obtained using a single mass bin is
easily sufficient for the current purpose of estimating the probability
of cluster membership.

NGC\,2516 shows an $RV$ distribution characteristic of a relaxed
cluster, centred at $23.6\pm0.1 $ km\,s$^{-1}$,
with an intrinsic dispersion $\le 0.79 \pm0.07$\,km\,s$^{-1}$. There
are 459 targets with a membership probability $>0.8$, about half of which have
$v\sin i >5$\,km\,s$^{-1}$. In the rest of the paper we conservatively 
consider this subset of stars where determining $R\sin i$
values. This sample should suffer a low level of contamination by
field stars, especially when we combine the $RV$ membership probabilities with the
likelihood of having $v \sin i > 5$ km\,s$^{-1}$ {\it and} the
likelihood of having a measurement of their rotation periods from
rotational modulation (see Sect.~3.1).  

NGC\,2547 shows an $RV$ distribution consistent with the presence of a
dominant cluster centered at $12.4\pm 0.1$\,km\,s$^{-1}$ with a
dispersion of $0.55\pm 0.09$\,km\,s$^{-1}$, together with a lesser,
more dispersed population with central $RV$ of $\sim
19.5$\,km\,s$^{-1}$. This is consistent with the results of Sacco et
al. (2015) who used a sub-sample of GES targets in NGC\,2547 selected
using the equivalent width of the Li~{\sc i}~6708\AA\ line to identify
the presence of a kinematically distinct, younger population in the
line of sight towards NGC\,2547, centred at $RV \sim
19$\,km\,s$^{-1}$. For our analysis we select only members of the main
cluster and find 156 targets with membership
probability $>0.8$, $\sim$75 per cent  of which have $v\sin i > 5$km\,s$^{-1}$.

NGC\,2264 shows a broader peak in its $RV$ distribution, with extended
tails containing a large fraction of fast rotating stars.  This is
consistent with NGC\,2264 being understood as a loose collection of
star-forming clumps rather than as a single, distinct, strongly bound
cluster (Tobin et al. 2015). This leads us to consider two membership samples, the
first where cluster members are those associated with the peak in the
$RV$ distribution at $19.8 \pm 0.2$\,km\,s$^{-1}$, giving 339 members
with membership probability $>0.8$, 
and a second where all 604 fast rotating stars
with $v \sin i >5$ km\,s$^{-1}$ and $-20<RV<60$\,km\,s$^{-1}$ are
considered to be potential cluster members.

\subsection{Comparison of $v\sin i$ with other work}
Both the precision and accuracy of the $v \sin i$ measurements are
important in estimating the $R \sin i$ distributions for stars in these clusters.
The precision of the $v\sin i$ measurements is estimated from an
empirical analysis of repeat measurements of $v\sin i$ recorded as part
of the GES project (Jackson et al. 2015). 
However, such measurements give no indication of the
absolute accuracy of the measurements, which is a combination of the
uncertainty in measurement and the uncertainty in absolute
calibration.

\begin{figure}
	\centering
		\includegraphics[width = 85mm]{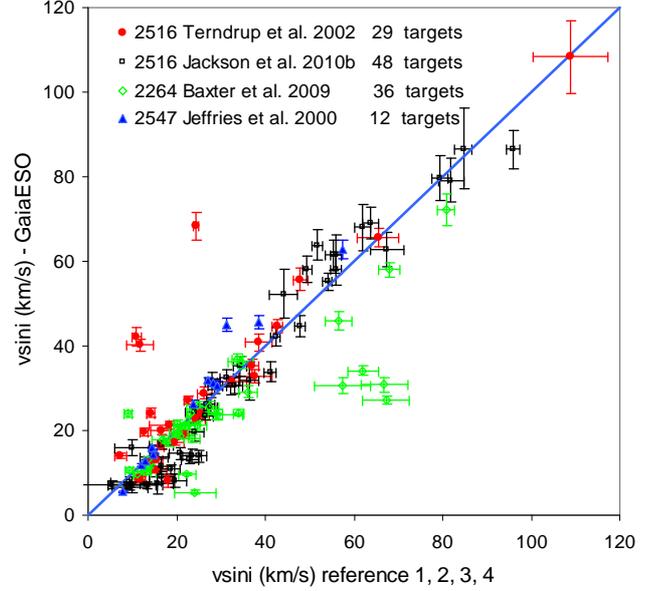}
\caption{Comparison of GES measurements of $v\sin i$ in NGC\,2516, NGC\,2547 and NGC\,2264 with previously reported values. The insert
shows the numbers of targets per cluster where matches were found with reported data.}
\label{fig3}
\end{figure}

To assess the calibration uncertainty, GES $v\sin i$ measurements were
compared with previously reported measurements for the same targets in
NGC\,2516 (Terndrup et al. 1998; Jackson \& Jeffries 2010b), NGC\,2547
(Jeffries, Totten \& James 2000) and NGC\,2264 (Baxter et
al. 2009). Figure 3 compares measured and reported values of $v\sin i$
for 125 targets where both sources showed $v\sin
i >5$\,km\,s$^{-1}$. The uncertainties shown for the GES data are
$\Delta_{v\sin i}$ (see Sect.~2.2). The uncertainties in the reported values,
$\sigma_{v\sin i}$ are taken from the source papers.

There is a good correlation between the two data
sets after accounting for the measurement
precisions. However, there are a number of outliers where
the difference between the two estimates of $v\sin i$ 
is much greater than the expected uncertainty in the difference
of the two values.  To some extent
this is expected since, at least for the GES data the measurement
uncertainty is described by a Student-t distribution (with $\nu$=2)
rather than a Gaussian distribution. To eliminate these outliers the
comparative data is clipped such that the discrepancy between the two
measurements must be less than 5 times its uncertainty, which is equivalent to selecting data between the
8th and 92nd percentiles of a Student's t-distribution with
$\nu=2$. Linear regression of the
clipped data then yields the following relation between the $v\sin i$
reported in the GES database and previously the reported values ($v\sin
i_{ref}$):
\begin{equation}
v\sin i = [-0.71\pm 0.64] + [1.01\pm 0.02]  v\sin i_{ref}\, .
\end{equation}
Thus there is no significant systematic difference between GES values of
$v\sin i$ and those previously reported for the same targets. This
calibration uncertainty is small compared to the
expected random uncertainty for individual targets --
usually $\ge 5$ per cent unless a target is observed on multiple
occasions.

\begin{figure*}
\centering
\includegraphics [width = 180mm]{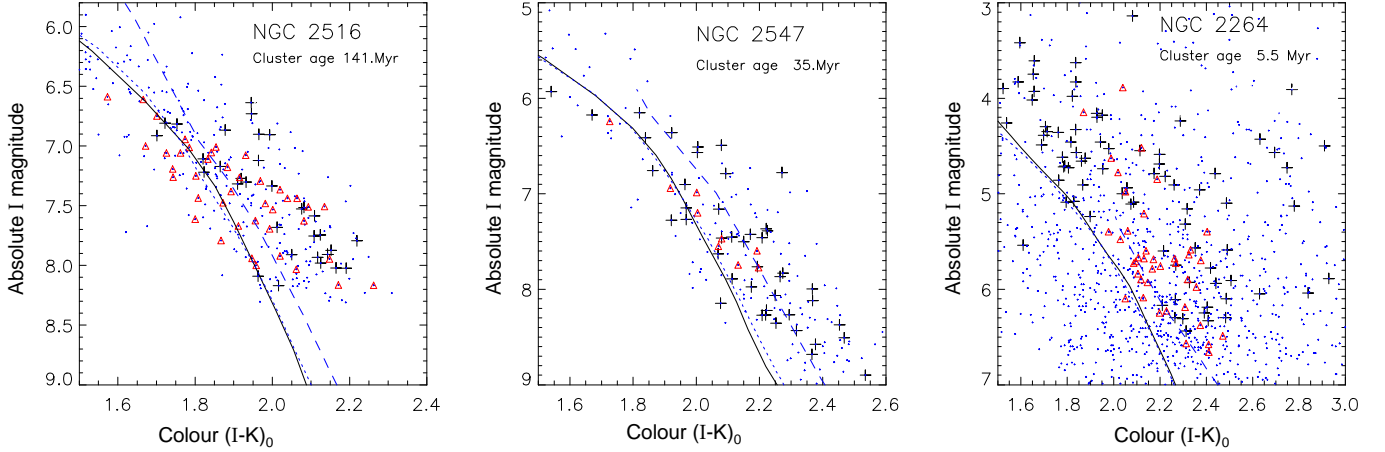}
\caption{Absolute $I$-magnitude versus intrinsic $I-K$ for 
GES targets in the open clusters
  NGC\,2516, NGC\,2547 and NGC\,2264 using the cluster parameters shown in
  Table 2. Points show all
  GES targets (with spectra that have $S/N>5$). Black crosses show stars
  with reported periods that are cluster members based on their $RV$ (see
  Sect.~2.3) with $v\sin i >5$\,km\,s$^{-1}$, red triangles show other GES 
  targets with reported periods. Lines show solar metallicity model
  isochrones interpolated to the cluster age for different
  evolutionary models, see Sect.~4.1, solid line -- BHAC15 (Baraffe et al. 2015),
  blue dotted line -- Dartmouth (Dotter et al. 2008), blue dashed line -- , Dartmouth modified 
  for magnetic fields (Feiden \& Chaboyer 2013).}
\label{fig4}
\end{figure*}

\section{Mean radii of cluster members}
\subsection{Targets with measured rotation period}

Projected radii, $R\sin i$, are calculated for the samples of probable members defined
in Sect.~2.4 that also have $v\sin i>5$\ km\,s$^{-1}$ and a reported
rotation period in the literature. We also demand a high quality
$v \sin i$ measurement, with  $\Delta_{v\sin i} < 0.2$. 
 
All three clusters have been the subject of photometric surveys that
attempted to measure the rotation periods of low mass cluster members
via the
rotational modulation caused by magnetic activity and starspots. Irwin
et al. (2007) found periods for 362 stars in NGC\,2516 over the mass
range (estimated from $M_K$ using the BHAC15 models) $0.15 <
M/M_{\odot} <0.7$. Cross correlation with the GES target list showed 77
matches, 32 of which are both members (based on their $RV$) and have
$v\sin i >5$ km\,s$^{-1}$ and $\Delta_{v\sin i}<0.2$.  Irwin et al. (2008) found rotation periods
for 176 stars in NGC\,2547 over the mass range $0.1<M/M_{\odot}<0.9$,
84 were matched with GES targets and 45 of these are $RV$ members with high quality $v
\sin i$ measurements. Makidon et al. (2004), Lamm et al. (2004) and Affer et al. (2013)
measured rotational periods in NGC\,2264 over the approximate mass range
$0.1<M/M_{\odot}<2.5$, yielding 226 matches with GES targets, 90 of
which are $RV$ members with high quality $v\sin i$ measurements (or 149 if we
take the broader definition of cluster membership in Sect.~2.4). 

Figure 4 shows $M_I$ vs $(I-K)_0$ colour magnitude plots\footnote{This
  colour-magnitude diagram was chosen as $I$ and $K$ magnitudes were
  available for all targets in all three clusters.} of the GES targets for
each cluster, highlighting targets for which $R \sin i$ has been calculated.  Also shown are
solar metallicity isochrones interpolated according to the logarithm of
cluster age for three different evolutionary models. The BHAC15 models
(Baraffe et al. 2015), 
Dartmouth models (Dotter et al. 2008) and a modified version of the
Dartmouth code that includes magnetic fields
(Feiden \& Chaboyer 2013; Feiden, Jones \& Chaboyer 2015, see Sect.~4.1). 
The $(I-K)$ colour of targets in NGC\,2516 and NGC\,2547
are somewhat redder than the non-magnetic BHAC15/Dartmouth
isochrones and are possibly better represented by the magnetic Dartmouth
models. However, this may also partly be due to unresolved binaries
  in the sample, or could simply reflect the fact that the BHAC15 models do
  appear to predict $I-K$ colours that are $\sim 0.2$ too blue at $M_{I} \sim 8$
  (see fig.5 in BHAC15). 
Results for NGC\,2264 show much more scatter in colour vs
magnitude for both the GES targets and the subset of members with
measured rotation period. This could be due to the presence of
circumstellar material, accretion, differential reddening or perhaps 
indicate some variation in the age of cluster members. 

Filled histograms in Fig.~1 show cluster members with resolved
values of $R\sin i$, i.e. those targets which are cluster members with
a measured rotation period, $v\sin i > 5$\,km\,s$^{-1}$ and $\Delta_{v\sin i}<0.2$. 
The median $v\sin i$ of this subset is 30 km\,s$^{-1}$  for NGC\,2516, 
20 km\,s$^{-1}$ for NGC\,2547 and 21 km\,s$^{-1}$ for NGC\,2264.
Table 3 lists the measured and calculated properties of all valid targets 
with $S/N > 5$ and measured rotation period. 

\begin{table*}
	\caption{ Measured and calculated properties of cluster targets
        with resolved $v \sin i$ and 
        a reported rotation period. The full Table~3 is available at the CDS.}
    \begin{tabular}{llllllllllllll}\hline\hline

Cluster &RA & Dec & $RV$ &$\Delta_{RV}$ &$v\sin i$ &$\Delta_{v\sin i}$& M$_K$ & $\log L/L_{\odot}$ &Membership&Period &$R\sin i$ &$\Delta_{R\sin i}$ &Flag\\
     &degrees &degrees& km\,s$^{-1}$ & km\,s$^{-1}$ & km\,s$^{-1}$ &   
&mag&               &probability&days& \multicolumn{2}{c}{$R_{\odot}$} & \\\hline

NGC\,2547 &122.1693 &-48.8938 &13.38 &0.81 &18.30 &0.14 &6.20 &-1.72 &0.94 &1.06 &0.39 &0.05&\\
NGC\,2547 &122.4060 &-49.0341 &12.48 &0.41 &15.10 &0.09 &5.30 &-1.31 &0.98 &1.96 &0.59 &0.05&\\
NGC\,2547 &122.2800 &-49.3207 &12.09 &0.18 & 5.50 &0.06 &4.15 &-0.68&0.98 &5.26 &0.58 &0.04&\\
NGC\,2547 &122.3004 &-49.4583 &13.03 &0.79 &17.90 &0.15 &6.32 &-1.77&0.96 &1.57 &0.56 &0.08&\\
NGC\,2547 &122.3077 &-49.5550 &16.85 &0.36 &16.10 &0.07 &5.31 &-1.36&0.16 &1.69 &0.55 &0.04&\\

    \hline
    \multicolumn {13} {l} {Note $R\sin i$ values are shown only where $v\sin i >5$ km\,s$^{-1}$ and
      $\Delta_{v\sin i} < 0.2$ where $\Delta_{v\sin i}$ is the \textit{relative} uncertainty in $v\sin i$}\\
    \multicolumn {13} {l} {Flag is set to 1 for targets in cluster NGC\,2264 where a (H-K) versus (J-H) colour colour plot indicates possible infra-red excess.}\\
      
  \end{tabular}
\label{tab3}
\end{table*}        

\subsection{Averaged radii as a function of luminosity}

The averaged radii of stars is calculated from $P$ and $v\sin i$ 
following the method described by Jeffries (2007). The product of these
quantities gives the projected radii in solar units, $R\sin i=
0.02\,P\,v\sin i$, where $P$ is in days and $v\sin i$ is in
km\,s$^{-1}$. Assuming that the stellar spin axes are randomly oriented
(e.g. Jackson \& Jeffries 2010a) then in principle the average radius can be found
by dividing the average $R\sin i$ for a group of similar stars by an average value of
$\sin i$. $R\sin i$ estimates for cluster members  
 are divided into $K$ magnitude (and later luminosity) bins with approximately
equal numbers of targets per bin. The radius value for each bin,
$\overline{R}$ is calculated from the {\it median} value of $R\sin i$
per bin, which is then corrected for the inclinations based on a distribution of $\sin i$ 
values and measurement uncertainties. Taking the median value of $R\sin i$ is preferred 
over the mean, since it minimises the effects of the expected extended tail in the distribution of
$v \sin i$ uncertainties.

A Monte Carlo method was used to determine the correction to the median
$R \sin i$ and the uncertainty in this correction. Samples of $N$
individual $R\sin i$ values, with estimates of measurement precision, and a known
probability distribution of $\sin i$ were simulated, under the
assumption that the
uncertainty in $R\sin i$ is dominated by $\Delta_{v\sin i}$ (see Section
2.1 and Table 3). It is further assumed that the
stellar spin axes are randomly distributed but that $R\sin i$ can only
be resolved if $\sin i > \tau$. The reason for this threshold is that 
stars with low inclinations do not exhibit sufficient
rotational modulation to enable a rotation period determination or do
not have sufficient equatorial velocity to yield a resolvable $v \sin i$. 

Random values of $\sin i$ were drawn from the distribution $P(i) = \sin
i/\cos(\arcsin \tau)$ where $\tau<\sin i<1$. The value of $\tau$ was
estimated directly from the measured distribution of $R\sin i$ values
about the median value of $R$ in two absolute magnitude bins for each
cluster (see Jackson et al. 2009 for details). For
the present data sets we find an average $\tau=0.16\pm0.11$, 
which corresponds to  $\overline{\sin i}=0.80\pm0.02$. (Note
that the effect of $\tau$ is small, because relatively few stars have a
low value of $\sin i$ in a random distribution of orientations.)
Multiple realisations are modelled using the appropriate uncertainties
for the dataset under consideration. The distribution of median values
is then analysed to determine the value of $\overline{R}$ and its
uncertainty.

\begin{table}
\centering
 	\caption{Averaged values of stellar radii NGC\,2516, NGC\,2547 
 	        and NGC\,2264 calculated from
          the product of GES measurements of $v\sin i$ and reported
          rotational periods, binned according to apparent $K$
          magnitude. }

\begin{tabular}{llllll}\hline\hline

Sample & $N$ & $K$ & $\log L/L_{\odot}$ & $M/M_{\odot}$ & $\overline{R}/R_{\odot}$ \\\hline
NGC\,2516 (a) & 15 & 13.15$\pm$05 &-1.19&0.60&0.62$\pm$0.08\\
              & 16 & 13.72$\pm$05 &-1.47&0.49&0.49$\pm$0.06\\

NGC\,2547 (a) & 15 & 12.51$\pm$08 &-1.01&0.66&0.73$\pm$0.07 \\
              & 15 & 13.29$\pm$05 &-1.36&0.44&0.78$\pm$0.11 \\
              & 14 & 13.89$\pm$05 &-1.63&0.29&0.57$\pm$0.07 \\

NGC\,2264 (a) & 21 & 11.62$\pm$04 &+0.01&1.22&2.00$\pm$ 0.16 \\
              & 20 & 12.09$\pm$02 &-0.18&1.04&1.76$\pm$ 0.16 \\
              & 20 & 12.47$\pm$03 &-0.35&0.82&1.67$\pm$ 0.15 \\
              & 20 & 13.16$\pm$06 &-0.68&0.52&1.21$\pm$ 0.12 \\

NGC\,2264 (b) & 33&  11.62$\pm$04 &-0.01&1.25&2.25$\pm$ 0.23\\
              & 34&  12.09$\pm$02 &-0.20&1.04&1.65$\pm$ 0.11\\
              & 31&  12.47$\pm$03 &-0.38&0.82&1.68$\pm$ 0.13\\
              & 23&  13.16$\pm$06 &-0.72&0.51&1.27$\pm$ 0.10\\

     \hline
   \multicolumn {6} {l} {(a) Averaged radii for members selected on $RV$ with }\\ 
   \multicolumn {6} {l} {\,\,\,\,\,\,\,\, period, $v\sin i > 5$\,km\,s$^{-1}$ and $\Delta_{v\sin i} <0.2$.} \\
   \multicolumn {6} {l} {(b) Averaged radii for all targets with period,}\\
   \multicolumn {6} {l} { \,\,\,\,\,\,\,\, $v\sin i >5$\,km\,s$^{-1}$ and $\Delta_{v\sin i} <0.2$.}\\

   \end{tabular}
 \label{tab:tab3}
\end{table}

Table 4 shows the average radii, $\overline{R}$, derived from the
$R\sin i$ estimates for members of each cluster with
  luminosities corresponding to masses in the range $0.2<M/M_{\odot}<1.4$ according to the BHAC15 model of Baraffe et al. (2015). The results are binned
according to $K$ magnitude with 15 to 20 targets per bin 
(or $\sim 33$ if the wider membership criteria is adopted for NGC~2264) . In NGC\,2547 and NGC\,2516 
 a single outlier with an exceptionally small $K$ mag was excluded. The 
relationships between $\overline{R}$ and apparent $K$ magnitude are
model- and distance-independent and not affected by the estimated
reddening to the cluster. Also shown in Table 4 are the model-, distance-
and reddening-dependent average luminosity and mass per bin derived from BHAC15 model
isochrones (Baraffe et al. 2015) using the cluster parameters shown in Table~2.

The over-radius of a set of targets relative to an evolutionary model
 is calculated as the median value of the over-radii  of  individual 
targets relative  to the model radii at equivalent luminosity, corrected for 
the effects  of inclination and measurement uncertainties using the same Monte-Carlo model.

\subsection{Measurement accuracy}

The uncertainty in $\overline{R}$ and in over-radius depends on the
number of targets. A simple estimate of the expected
uncertainty is given by considering the ideal case where the
uncertainties in $P$ and $v\sin i$ are small compared to the uncertainty
in the mean value of $\sin i$, which (in the absence of a lower cut off)
has a mean value of $\pi/4 \pm 0.22/\sqrt{N}$ where $N$ is the 
number of targets per bin. Hence the minimum uncertainty in $\overline{R}$
for $N=15$ is $\sim 7$ per cent, and would reduce to
3 per cent for $N=100$. The estimates of uncertainty in $\overline{R}$
shown in Table 4, from Monte Carlo simulations, 
also take account of the measurement errors in $v\sin i$
and the non Gaussian distribution of $\sin i$ (see Sect.~3.2), giving
average uncertainties of $\sim 12$ per cent for our bin occupancies of $14 < N <
34$.
%sqrt(2/3.0-(!dpi/4.0)^2

The procedure described in Sect.~3.2 gives an unbiased estimate of
$\overline{R}$ provided that $P$ and $v\sin i$ are
themselves unbiased and the distribution of $\sin i$ is modelled
correctly. Hartman et al. (2010) pointed out that differential
rotation may systematically increase the measured values of $P$. However
the rate of differential rotation is low for active K- and M- dwarfs
(e.g. Reinhold, Reiners \& Basri 2013), and the effect of any positive bias
in $P$ produced by differential rotation is offset by a corresponding
negative bias in measured $v\sin i$. The net changes in inferred radii using the
method discussed here should be less than 1 per cent (see
discussion in Jackson \& Jeffries 2014b). A second source of bias is
uncertainty in the distribution of $\sin i$, in particular the effect
of uncertainty in the lower cut off value, $\tau$, below which $R\sin i$
values cannot be obtained.  However, the large uncertainty in $\tau$
 given in Sect~3.2 corresponds to only a 2 per cent systematic error in 
 $\overline{\sin i}$ and $\overline{R}$ and so is much less important
than the statistical uncertainties.

A further source of uncertainty could be contributed by targets
  that are unresolved spectroscopic
  binaries. For some fraction of these targets, the measured values of $v\sin i$
  could be systematically higher than the true $v \sin i$ of the primary
  star, depending on the difference in RVs of the primary and secondary
  stars and their relative contribution to the GES spectra. This would
  lead to an upward bias in the estimate of $R \sin i$. Appendix 1
  presents an analysis of the effects of binarity on the measured value
  of $v\sin i$ which is then used to estimate the increase in $R\sin i$
  for individual observations assuming the binary distribution
  described in Sec.~2.4. Calculations of the averaged radii (Sec.~3.2)
  were repeated with a correction made for this binary bias in $R\sin
  i$.  For NGC\,2516 and NGC\,2547 there was a small reduction ($<2$
  per cent) in the averaged radii whilst for NGC\,2264 the downward
  correction was higher at $\sim 4$ per cent, due to its more massive
  members that have a higher assumed binary fraction. The bias caused
  by binarity is therefore not completely negligible, but is still
  small compared to the statistical uncertainties for the current data
  set.

\section{Discussion}

\begin{figure*}
\centering
\includegraphics [width = 185mm]{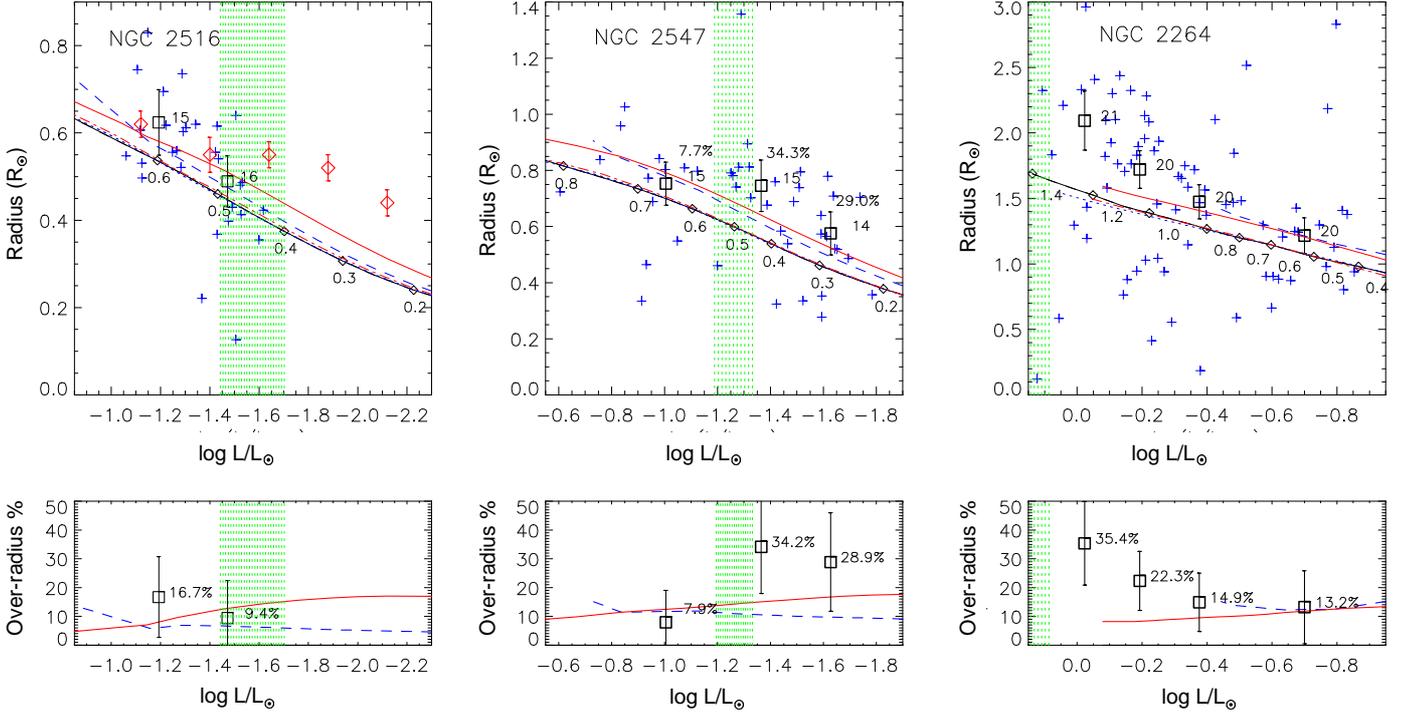}
\caption{Stellar radius versus luminosity for stars in the open
  clusters NGC\,2516, NGC\,2547, NGC\,2264. Crosses in the upper plots
  show the $R\sin i$ of individual targets normalised by the average
  value of $\sin i$ (see Sect.~3.2). Squares with error bars show the averaged
  radii in bins of luminosity; adjacent numbers indicating the
  numbers of targets per bin. Diamonds with error bars show averaged
  radii (calculated in a similar way) from 
   an alternate data set in NGC\,2516 from Jackson et al.
  (2009). Lines show solar metallicity model isochrones of $R$ versus
  $\log L$ interpolated to the cluster ages in Table~2 from several evolutionary
  models (see Sect.~4.1): the black solid line (with circles indicating fiducial mass
  points) -- BHAC15 (Baraffe et al. 2015), blue dotted line -- Dartmouth
  (Dotter et al. 2008), blue dashed -- Dartmouth modified for magnetic
  fields (Feiden \& Chaboyer 2013; Feiden et al.
  2015), red dot-dashed -- YREC, and red solid line -- YREC modified for an
  effective spot coverage of 30 per cent (Somers \&
  Pinsonneault 2015). The shaded area indicates the range of
  luminosities over which stars develop radiative cores according to
  the BHAC15 model. Stars to the left of this region have radiative
  cores, while stars to the right are fully convective. 
  The lower plots show the ``over-radius'', expressed
  as a percentage of the predicted radius from the BHAC15
  model. Dashed and solid lines show the over-radius, with respect
  to their non-magnetic counterparts, predicted by
  the  Dartmouth-magnetic model and the YREC with starspots
  model respectively.} 
\label{fig5}
\end{figure*}

\subsection{Comparison with model isochrones}
Figure 5 compares the measured radii, as a function of luminosity, with
the predictions of three current evolutionary models, two of which have
also been adapted to explicitly include the effects of magnetic fields and activity 
on the evolution of low-mass stars. The models considered are:
\begin{enumerate}
\item The solar metallicity isochrones of $R$
  versus $\log L$ for the BHAC15 model (Baraffe et al. 2015),
  interpolated to the cluster ages in Table~2. Open circles on the
  isochrones mark fiducial mass  points. 
\item The Dartmouth (Dotter et al. 2008) and Dartmouth-magnetic models
  (Feiden \& Chaboyer 2013; Feiden, Jones \& Chaboyer 2015).  The latter is a version of the
  Dartmouth evolutionary code modified to take account of the effect of
  magnetic fields on the equation of state and on mixing length theory
  in the stellar interior of magnetically active stars. The isochrones
  shown here correspond to the
  rotational dynamo model of Feiden \& Chaboyer (2013) with a surface field strength
  of 2.5 kG, as described in detail by Malo et al. (2014). 

\item The YREC and YREC-spot models, where the standard YREC
  evolutionary code (van Saders \& Pinsonneault 2013) has been adapted
  by Somers and Pinsonneault (2015;  isochrones supplied by Somers,
    private communication) to model the effect of dark
  starspots on the evolution of magnetically active, low-mass stars. 
  The specific isochrones in Fig.~5 show predicted radii for
  stars with 0 and 50 per cent coverage of starspots with an average
  spot temperature of 30 per cent of the temperature of the (unspotted)
  photosphere (equivalent to an effective coverage by dark spots of 
  30 per cent).
\end{enumerate}

In the absence of imposed magnetic fields and/or starspots all three
models show almost identical radii as a function of luminosity in the
upper panels of Fig.5. The effect of magnetic fields and/or starspots
is to increase radii at a given luminosity, although the percentage
increase in radius at a given field strength or spot coverage varies
with both mass and age. This over-radius, expressed with respect to
their non-magnetic counterpart models, is compared with our
measurements of average radii with respect to the BHAC15 isochrones in
the lower panels of Fig.~5. The shaded areas in Fig.~5 indicate the
range of luminosities over which stars are expected to develop
radiative cores according to the BHAC15 model at the assumed age of the
clusters. i.e. stars to the left of the shaded area are expected to
have radiative cores, stars to the right are expected to be fully
convective and still be descending their Hayashi tracks. The width of
these regions corresponds to $\pm 0.05M_{\odot}$, the resolution of the
model grids in our possession.

The percentage increase in radius shown by the Dartmouth-magnetic model
with respect to non-magnetic models is smaller for the low-mass, fully
convective stars. This is consistent with the results of Feiden \&
Chaboyer (2012, 2013) for MS stars, where magnetic inhibition of
convection can produce significant inflation for stars with radiative
cores but has less effect (for a given magnetic field strength)
for fully convective stars. Starspots have the opposite effect,
producing a larger radius inflation for fully-convective PMS
stars. Jackson \& Jeffries (2014b) used polytropic models to show that
starspots slow stellar contraction along Hayashi tracks, asymptotically
inflating the radii of PMS stars by a factor of $(1-\beta)^{-n}$
compared to unspotted stars of the same luminosity, where $\beta$ is
the equivalent covering fraction of dark starspots and $n \sim 0.5$
. This is a much stronger inflation than predicted by Spruit \& Weiss
(1986) for spotted MS stars with the same $\beta$. The over-radius
predicted for fully convective stars in NGC\,2516 and NGC\,2547 by the
YREC-spot model with $\beta=0.3$ (shown in Fig.~5) is $\simeq 15-18$ per
cent, which is fully consistent with the simple scaling derived from
polytropic models.

\subsection{NGC\,2547}
We consider this cluster first since we have data for both fully 
convective PMS stars and ZAMS stars with radiative cores. 
Figure 5 shows averaged radii for 44 targets divided into 3 bins, one
to the left of the shaded area containing stars with radiative cores
and two to the right, containing fully convective stars on their Hayashi
tracks. The average over-radius for the full sample, relative to the
predictions of the BHAC15 model is $18\pm 7$ per cent. This result
applies to cluster members having $>80$ per cent probability of
membership according to their $RV$ (see Sect.~2.4). There are in
addition 15 targets with measured $R\sin i$ but $<80$ per cent membership
probability (see Table 3). This latter sample is expected to contain a
significant proportion of stars from a secondary population identified
by Sacco et al. (2015) as kinematically distinct and much younger than
NGC\,2547. Taking an age of $\sim 10$\,Myr (Sacco et al. 2015) and
assuming $R \propto t^{-1/3}$ then the radius of any stars belonging to
this younger population should be $\sim 60$ per cent larger than
members of NGC\,2547. The mean over-radius of these 15 targets is
$45\pm 21$ per cent and thus consistent with a
significant proportion belonging to the younger
population identified by Sacco et al.

For NGC\,2547 the highest luminosity bin in Fig.~5, containing stars with radiative cores,
has $\overline{R}$ compatible (within $\sim 1\sigma$)
with both magnetic and non-magnetic evolutionary models. The low-luminosity
bins containing fully convective stars show significantly higher radii
than predicted by the non-magnetic evolutionary models. The average over-radius for
data in these two lower mass bins is $29\pm 10$ per cent compared to the
BHAC15 isochrone. The possible systematic errors discussed in section
3.2 associated with differential rotation, uncertainties in the
threshold inclination for $R\sin i$ measurement {\it or binarity}, are small compared
with the statistical uncertainties. There are additional systematic
uncertainties due to possible errors in the assumed distance modulus,
reddening and age. These
effects are also small for NGC\,2547 where the combined uncertainty in
$(m-M)_0$ and $E(B-V)$ leads to an uncertainty of only 0.04 dex in estimated
$\log L$ values, which
corresponds to less than a 2.5 per cent change in radius at a fixed
luminosity. A change in age of $\pm 3$ Myr leads to a change of 
only $\sim \pm 1$ per cent in the average over-radius with respect to 
the BHAC15 models. It follows that the estimated over-radius (with
respect to non-magnetic models) for convective stars 
is significantly larger ($>2 \sigma$) than both the random errors in the
measurements and the systematic uncertainties in the method. 

The observed over-radius of $29\pm 10$ per cent could be caused by either 
magnetic inhibition of convection, by starspots or by a combination of the two. 
\begin{itemize}
\item The curve in the lower panel of Fig.~5 shows the over-radius
  produced by Dartmouth-magnetic models for a rotational dynamo and
  surface magnetic field strength of 2.5\,kG. The over-radius amounts
  to about 10 per cent for the low-mass fully-convective stars. The
  plots in Feiden et al. (2015) suggest that a much larger surface
  magnetic field would be required to explain a $\sim$30 per cent
  over-radius, but this would be incompatible with the level of
  (integrated) magnetic field measured on rapidly rotating M-dwarfs of
  3$\pm$1\,kG (Reiners, Basri \& Browning 2009). The turbulent dynamo
  model of Feiden \& Chaboyer (2013) has a much smaller effect on
  low-mass PMS stars (e.g. see Malo et al. 2014).

\item The YREC-spot models with $\beta=0.3$ predict an over-radius of
  about 15-18 per cent for fully convective stars in NGC\,2547, which is
  almost compatible with the observations. Using the scaling relation
  between over-radius and spot coverage of Jackson \& Jeffries (2014b),
  the measured over-radius would actually correspond to $\beta =
  0.43^{+0.08}_{-0.11}$. If we adopt the temperature ratio of $0.70\pm
  0.05$ between spots and unspotted photosphere, as measured by O'Neal et
  al. (2004) and O'Neal (2006) in some active G/K dwarfs, this would require
  a spot area coverage of $57^{+11}_{14}$ per cent. If the temperature
  ratio was larger, as advocated by Feiden \& Chaboyer (2014), then
  this would require an even higher spot coverage.

\item Alternatively, the observed level of radius inflation could result
  from the combined effects of inhibition of convection in the stellar
  interior and inhibition of radiative heat transfer from the stellar
  surface due to starspots. This scenario was considered as one of
  several options to account for the over-radius of PMS stars in NGC\,2516
  by MacDonald \& Mullan (2013). If we were to assume that the effects were
  additive, then a combination of the 2.5~kG
  Dartmouth-magnetic model and starspots with $\beta=0.32$ would give
  an over-radius of 30 per cent in fully convective stars
  (coincidentally this is almost exactly the sum of the over-radii shown
  for the two magnetic models in the lower panel of Fig.~5).
\end{itemize}

\subsection{NGC\,2516}

GES targets in NGC\,2516 are restricted to $I < 16.5$. Consequently
there are relatively few targets with measured periods from the
photometric survey of Irwin et al. (2007), which covered stars with
$14<I<18$. Only a fraction of these are relatively fast rotators ($v\sin i >
5$\,km\,s$^{-1}$), resulting in only 32 targets with measured $R\sin
i$. In Fig.~5 these data are divided into two bins both of which lie
leftward of the shaded area indicating stars with radiative
cores. According to the BHAC15 evolutionary model, stars of this
luminosity and with an age of $\sim$140\,Myr should be ZAMS stars. An
average over-radius of 14$\pm9$ per cent relative to the BHAC15 model is
inferred from the $R\sin i$ data. This falls to $\sim$8~per cent
relative to the Dartmouth-magnetic/YREC-spot models. The measurement
uncertainties are therefore too large to discern whether the magnetic
or non-magnetic models better describe the data.

Also shown in Fig.~5 are estimates of $\overline{R}$ from previous work
(Table 1 of Jackson \& Jeffries 2014b), which also used Giraffe
spectroscopy to estimate $v \sin i$ and periods from Irwin et
al. (2007), to determine $R \sin i$ for a sample that extended to
lower luminosities. The inferred mean radii from both datasets show
reasonable agreement for ZAMS stars over a relatively narrow common mass range
($0.45 < M/M_{\odot}<0.65$) and these are consistent with either
magnetic or non-magnetic models.  The results of Jackson \& Jeffries
(2014b) however demonstrate a much larger over-radius ($\sim 40$ per cent) for fully
convective PMS stars.  This is the same situation as found in
NGC\,2547 and could be explained in the same way. Unfortunately no
direct comparison for PMS stars in NGC\,2516 can be made using the GES
data.

\subsection{NGC\,2264}
Figure 5 shows average radii for targets in the mass range
 $0.4 <M/M_{\odot} <1.4$ according to the BHAC15 model of Baraffe et al. (2015).
All but two targets show luminosities to the right of the shaded area indicating
that they are PMS stars still on their Hayashi tracks. 
Table 4 lists the average radii for two cases; the first calculated using 
$R\sin i$ values for targets identified as $RV$ cluster members in section
2.2, the second using a sample of all targets with measured $P$ and
$v\sin i>5$\,km\,s$^{-1}$ and $-20 < RV < 60$ \,km\,s$^{-1}$. The broader sample 
generally shows higher radii but not at a significant level. The results, 
 as measured, are consistent with the majority of the faster rotating 
targets being members of a single young cluster. 

The colour-magnitude plot (Fig.~4) shows a number of 
targets with measured $R\sin i$ that have $(I-K)_0 >0.5$ 
mag redder than the BHAC15 isochrone, too red to be unresolved binaries
and suggesting that a fraction of the targets in NGC\,2264 show 
excessive reddening and/or 
an infra-red excess due to the presence of disks, which could
in turn effect their estimated luminosities and consequently our
estimates of over-radius. To test this hypothesis, targets with measured
period were plotted on a $(J-H)$ versus $(H-K)$ colour colour diagram and 
those showing higher than expected $(H-K)$ colour with respect to the
usual reddening band were flagged in Table 3 as having 
a possible infra-red excess. Excluding these flagged targets reduces
the sample size by 
10--20 per cent per bin but makes no makes no measurable difference
($<2$ per cent) to the averaged
radii and over-radii. Hence, including the few stars with overt signs
of circumstellar material appears not to 
significantly affect the results. 

Comparison of the measured radii with the 5.5\,Myr isochrone of
  the BHAC15 model shows an average over-radius of $19 \pm 5$ per
  cent. Lower mass stars ($<1M_{\odot}$) show an average over-radius
  similar to that predicted by either the Dartmouth-magnetic model or
  the YREC-spot model.  However, for NGC\,2264 the inferred over-radii
  depends strongly on the adopted cluster age and to a lesser extent on
  its distance. If we assume an
  uncertainty of $\pm 2.5$\,Myr in age at a fixed distance, 
  which is comparable to the age
  spread suggested by Sung et al. (2010), then the over-radius varies
  from $8 \pm 5$ per cent for an age of 3\,Myr to $27 \pm 5$ per cent
  at 8\,Myr. In principle, comparison of measured and predicted radii
could be used to help constrain the age of very young clusters
(e.g. Jeffries 2007).  However such comparisons are strongly model
dependent since the predicted radius at a given age depends on how the
effects of the rotationally induced magnetic fields are represented (if
at all) in the evolutionary model.

\section{Summary}

GES determinations of $RV$ for photometrically selected candidates have
been used to define samples of low-mass members in the three young
clusters NGC\,2516, NGC\,2547 and NGC\,2264.
By combining GES measurements of projected rotation velocity and
published rotation periods for a subset of these stars, 
we have estimated model-independent
projected radii as a function of $K$ magnitude for all three clusters. Average
radii as a function of luminosity are calculated and compared with isochrones 
predicted by several evolutionary models, including variants that
incorporate the influence of magnetic inhibition of convection 
or photospheric starspots. 

\begin{itemize}

\item The radii of these young, fast-rotating stars in NGC\,2547 and NGC\,2516 are, on
  average, larger than predicted by non-magnetic evolutionary
  models. However, the discrepancy is mostly apparent for the
  lower-mass, fully convective PMS stars. The current uncertainties in the age 
  and distance for NGC\,2264 preclude any meaningful comparison of
  stellar radii with the models.

\item The radii measured for higher-mass ZAMS stars in NGC\,2547 and NGC\,2516 are
  consistent with either non-magnetic or magnetic models. However,
  the difference between the predicted radii from these models ($\simeq
  10$ per cent) is
  comparable to the precision with which our limited samples permit
  the determination of average radii. A more decisive test would
  require an additional $\sim 100$ determinations of rotation periods
  to reduce the measurement uncertainties below 5 per cent.

\item In contrast the average radii of the lower-mass PMS stars
  significantly exceed those predicted by non-magnetic models -- by $29
  \pm 10$ per cent for members of NGC\,2547 and by $\sim 40$ per cent
  for members of NGC\,2516 studied previously using similar techniques. To explain
  this inflation with the
  magnetic models would either require: (i) rotational dynamos that
  produce very large
  surface magnetic fields ($>2.5$ kG), that may be incompatible with
  direct measurements of surface fields on active stars; (ii) starspots
  that block 30-50 per cent of the photospheric flux; or a more
  moderate combination of both.
\end{itemize}

\nocite{Affer2013a}
\nocite{Somers2014a}
\nocite{Meynet1993a}
\nocite{Jeffries1998a}
\nocite{Lyra2006a}
\nocite{Terndrup2002a}
\nocite{Jeffries2005a}
\nocite{Naylor2006a}
\nocite{Dzib2014a}
\nocite{Baxter2009a}
\nocite{Turner2012a}
\nocite{Sung1997a}
\nocite{Kamezaki2013a}
\nocite{Turner2012a}
\nocite{Sung2010a}
\nocite{Magrini2015a}
\nocite{Jackson2015a}
\nocite{Skrutskie2006a}
\nocite{Carpenter2001a}
\nocite{Baraffe2015a}
\nocite{Sacco2015a}
\nocite{Tobin2015a}
\nocite{Jackson2010a}
\nocite{Jeffries1998a}
\nocite{Jeffries2000a}
\nocite{Baxter2009a}
\nocite{Irwin2007a}
\nocite{Irwin2008a}
\nocite{Makidon2004a}
\nocite{Lamm2004a}
\nocite{Jackson2010b}
\nocite{Jeffries2014a}
\nocite{Pasquini2002a}
\nocite{gilmore2012a}
\nocite{Randich2013a}
\nocite{Jackson2009a}
\nocite{LopezMorales2007a}
\nocite{Morales2009a}
\nocite{Torres2010a}
\nocite{Mullan2001a}
\nocite{Chabrier2007a}
\nocite{Feiden2012a}
\nocite{Feiden2013a}
\nocite{Macdonald2012a}
\nocite{Boyajian2012a}
\nocite{Dotter2008a}
\nocite{Baraffe1998a}
\nocite{Reinhold2013a}
\nocite{Jackson2014a}
\nocite{Jackson2014b}
\nocite{Hartman2010a}
\nocite{Feiden2015a}
\nocite{Feiden2013a}
\nocite{Feiden2014a}
\nocite{Spruit1986a}
\nocite{Macdonald2013a}
\nocite{Jeffries2001a}
\nocite{Raghavan2010a}
\nocite{Jeffries2007a}
\nocite{Somers2015a}
\nocite{vanSaders2013a}
\nocite{Malo2014a}
\nocite{Reiners2009a}
\nocite{ONeal2004a}
\nocite{ONeal2006a}
\nocite{Sung2008a}
\nocite{Jeffries2004a}
\nocite{Naylor2002a}
\nocite{Cottaar2012a}

\begin{acknowledgements}
RJJ wishes to thank the UK Science and Technology Facilities Council
for financial support.  Based on data products from observations made
with ESO Telescopes at the La Silla Paranal Observatory under programme
ID 188.B-3002. These data products have been processed by the Cambridge
Astronomy Survey Unit (CASU) at the Institute of Astronomy, University
of Cambridge, and by the FLAMES/UVES reduction team at
INAF/Osservatorio Astrofisico di Arcetri. These data have been obtained
from the Gaia-ESO Survey Data Archive, prepared and hosted by the Wide
Field Astronomy Unit, Institute for Astronomy, University of Edinburgh,
which is funded by the UK Science and Technology Facilities Council.
This work was partly supported by the European Union FP7 programme
through ERC grant number 320360 and by the Leverhulme Trust through
grant RPG-2012-541. We acknowledge the support from INAF and Ministero
dell' Istruzione, dell' Universit\`a' e della Ricerca (MIUR) in the
form of the grant "Premiale VLT 2012". The results presented here
benefit from discussions held during the Gaia-ESO workshops and
conferences supported by the ESF (European Science Foundation) through
the GREAT Research Network Programme.
\end{acknowledgements}

\bibliographystyle{aa}  
\bibliography{references}  

\appendix
\section{Estimating the effect of binarity on measured values of $v\sin i$}

In the Gaia-ESO survey pipeline $v\sin i$ is estimated from the broadening of
spectral lines produced by stellar rotation. In the case of unresolved
binaries, the line widths of the measured spectrum may be increased relative
to those of the primary star depending on the difference in RV between the
primary and secondary star and the relative contribution of the
secondary to the observed spectrum.

The additional broadening can be determined as a function of the
line of sight velocity of the primary relative to the centre of mass, $RV_a$,
and the relative flux contribution of the secondary at the wavelength of the
observed spectra, $f_b/f_a$, by measuring the
FWHM, $W$, of a Gaussian profile fitted to the sum of two separate
Gaussian profiles. The first Gaussian represents the primary star with central
velocity $RV+RV_a$ and FWHM $W_a$; the second Gaussian represents the
secondary with velocity $RV-RV_a/q$ and FWHM $W_b$, where $q$ is the
binary mass ratio.

To a reasonable approximation the effect of rotational broadening is to
increase the measure FWHM as $W=W_0\sqrt{1+(v\sin i/C)^2}$ where $W_0$
is the unbroadened line width, $C$ is a constant dependent on the
resolution, $R_{\lambda}$, of the spectrograph and the speed of light,
$c$, as $C=0.895c/R_{\lambda}$, giving $C=15.8$\,km\,s$^{-1}$ for the Giraffe
spectrograph using order sorting filter HR15N (see Jackson et al. 2015
for derivation). Using this expression for $W$, the apparent increase in
the measured rotational velocity relative to the $v\sin i$ of the
primary is given by, $\Delta v \sin i = C\sqrt{W^2/W_a^2-1}$ where $W_a
\approx (c/R_{\lambda})\sqrt{1+(v\sin i)^2/C^2}$.

\begin{figure}
	\centering
		\includegraphics[width = 85mm]{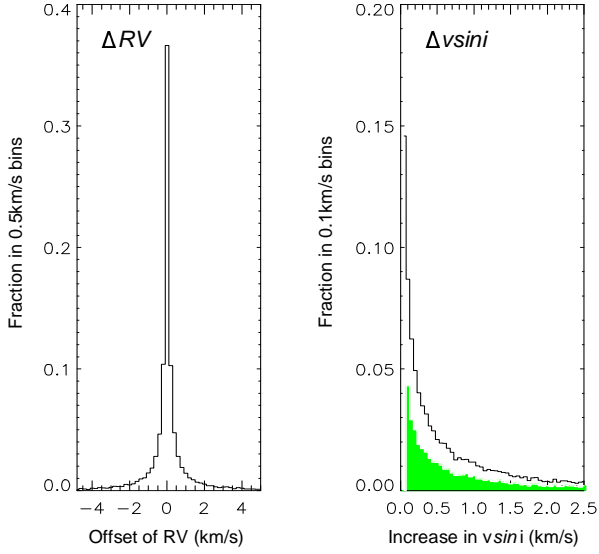}
\caption{The distribution of the offsets in measured values of $RV$ and of
  $v\sin i$ for a population of binary stars
   using the binary distribution parameters 
  described in
  section 2.4, a luminosity-mass relation of $L \propto
  M^{3}$ and a true $v\sin i = 20$ km\,s$^{-1}$. The black line shows
  the distribution assuming all stars are unresolved binaries. The
  solid histogram shows the actual probability distribution of $\Delta
  v\sin i >0$ for a random distribution of targets with an initial
  binary fraction of 0.5 and where targets showing an offset $\Delta RV >
  5$ km\,s$^{-1}$ are excluded from the distribution since they would not be
  classified as cluster members (see Sect.~2.4).}

\label{fig6}
\end{figure}

Fig.~A.1 shows the distribution of the offset in the measured recession
velocity, $\Delta RV$ (relative to the centre of mass) and the increase
in rotational velocity, $\Delta v\sin i$ obtained for a random
distribution of binary stars (as described in Sect. 2.3) with similar
true rotational velocities of $v\sin i=20$\,km\,s$^{-1}$, where the relative
flux from the secondary star is assumed to vary as a fixed power of
mass $f_b/f_a = q^n$ with $n \approx 3$. Binary pairs showing large
changes in $RV$ will mostly have a membership probability $< 0.8$
(e.g. see
Fig.~2). In this example binaries with $\Delta RV > 5$km\,s$^{-1}$ are
assumed to be non-members and are not included in the distribution of
$\Delta v\sin i$. Of course, only a fraction of stars are in binaries
and this further
reduces the proportion of stars that show large offsets in $v\sin i$ (see
Fig.~A.1).

This analysis is used to calculate the additional bias in the averaged
radius due to binarity. First the fractional increase in individual
$R\sin i$ values is calculated from the median increase in $1+ \Delta
v\sin i/v\sin i$ using the target $v\sin i$ and (mass-dependent) binary
fraction. The increased values of $R\sin i$ are then used to
recalculate the averaged values of $\overline{R}$ using the method
described in Sect~3.2.

%%%%%%%%%%%%%%%%%%%%%%%%%%%%%%%%%%%%%%%%%%%%%%%%%%%%%%%%%%%%%%%%%

\label{lastpage}
\end{document}